\documentclass[journal]{IEEEtran}

\usepackage{ifpdf}
\usepackage{cite}
\usepackage{amsmath,amssymb,amsfonts}
\usepackage{algorithmic}
\usepackage{graphicx, adjustbox}
\usepackage{array}
\usepackage{enumitem}
\usepackage{breqn}
\usepackage{ dsfont }
\usepackage{ upgreek }
\usepackage{textcomp}
\usepackage{multirow}
\usepackage{algorithm}
\usepackage{algorithmic}
\usepackage{caption}
\usepackage{pgfplots}
\usepackage{tkz-euclide}
\usepackage{subcaption}
\usepackage{tikz}{\tiny }
\usepackage{hyperref}
\usepackage{balance}
\begin{document}
	
\title{Intelligent Blockchain-based Edge Computing 
		\\ via Deep Reinforcement Learning: \\
		Solutions and Challenges}
	
	\author{\IEEEauthorblockN{ Dinh C. Nguyen, Van-Dinh Nguyen, Ming Ding, Symeon Chatzinotas, Pubudu N. Pathirana, \\ Aruna Seneviratne, Octavia Dobre,~\IEEEmembership{Fellow,~IEEE}, and Albert Y. Zomaya,~\IEEEmembership{Fellow,~IEEE} }
		
	\thanks{Dinh C. Nguyen is with the School of Electrical and Computer Engineering, Purdue University, USA.}
	
	\thanks{Van-Dinh Nguyen and Symeon Chatzinotas are with the Interdisciplinary Centre for Security, Reliability and Trust (SnT), University of Luxembourg.}
	
	\thanks{Ming Ding is with Data61, CSIRO, Australia.}
	
	\thanks{Pubudu N. Pathirana is with the School of Engineering, Deakin University, Australia.}
	
	\thanks{Aruna Seneviratne is with School of Electrical Engineering and Telecommunications, UNSW, Australia.}
	
	
	\thanks{Octavia Dobre is with the Faculty of Engineering and Applied Science, Memorial University, Canada.}
	\thanks{Albert Y. Zomaya is with the School of Computer Science, The University of Sydney, Australia.}
}
	
	\markboth{Accepted at IEEE Network Magazine}%
	{}

	\maketitle
	
	\begin{abstract}
The convergence of mobile edge computing (MEC) and blockchain is transforming the current computing services in wireless Internet-of-Things networks, by enabling task offloading with security enhancement based on blockchain mining. Yet the existing approaches for these enabling technologies are isolated, providing only tailored solutions for specific services and scenarios. To fill this gap, we propose a novel cooperative task offloading and blockchain mining (TOBM) scheme for a blockchain-based MEC system, where each edge device not only handles computation tasks but also deals with block mining for improving  system utility. To address the latency issues caused by the blockchain operation in MEC, we develop a new Proof-of-Reputation consensus mechanism based on a lightweight block verification strategy. To accommodate the highly dynamic environment and high-dimensional system state space, we apply a novel distributed deep reinforcement learning-based approach by using a multi-agent deep deterministic policy gradient algorithm. Experimental results demonstrate the superior performance of the proposed TOBM scheme in terms of enhanced system reward, improved offloading utility with lower blockchain mining latency, and better system utility, compared to the existing cooperative and non-cooperative schemes. {The paper concludes with key technical challenges and possible directions for future blockchain-based MEC research.} 
	\end{abstract}
	
	\IEEEpeerreviewmaketitle

\section{Introduction}
Recent advances in wireless Internet-of-Things (IoT) have promoted the proliferation of  mission-critical applications, e.g., augmented reality and autonomous driving, which rely heavily on edge devices (EDs) to collect data from IoT sensors to serve end users. To meet the ever-growing computation demands of EDs, mobile edge computing (MEC) has been proposed as a promising technique to improve the computation experience of EDs, by offloading computationally-intensive IoT tasks to a nearby MEC server located at a base station (BS) \cite{1}. Multiple EDs can share computation and communication resources of the BS to handle data tasks without device's battery depletion. Task offloading with MEC thus becomes a viable solution to satisfy various EDs' computation demands, {thus enhancing} the \textcolor{black}{quality-of-experience} (QoE) of end users. Furthermore, to provide security in MEC systems, blockchain \cite{3} has emerged as a strong candidate due to its decentralization, immutability, and traceability, which forms blockchain-based MEC (B-MEC) paradigms \cite{4}. Also, blockchain can build {trusted B-MEC schemes} by employing community verification among network entities (e.g., EDs) via mining mechanisms such as Delegated Proof of Stake (DPoS) \cite{add1} without requiring a central authority.
\begin{figure}
	\centering
	\includegraphics[width=0.95\linewidth]{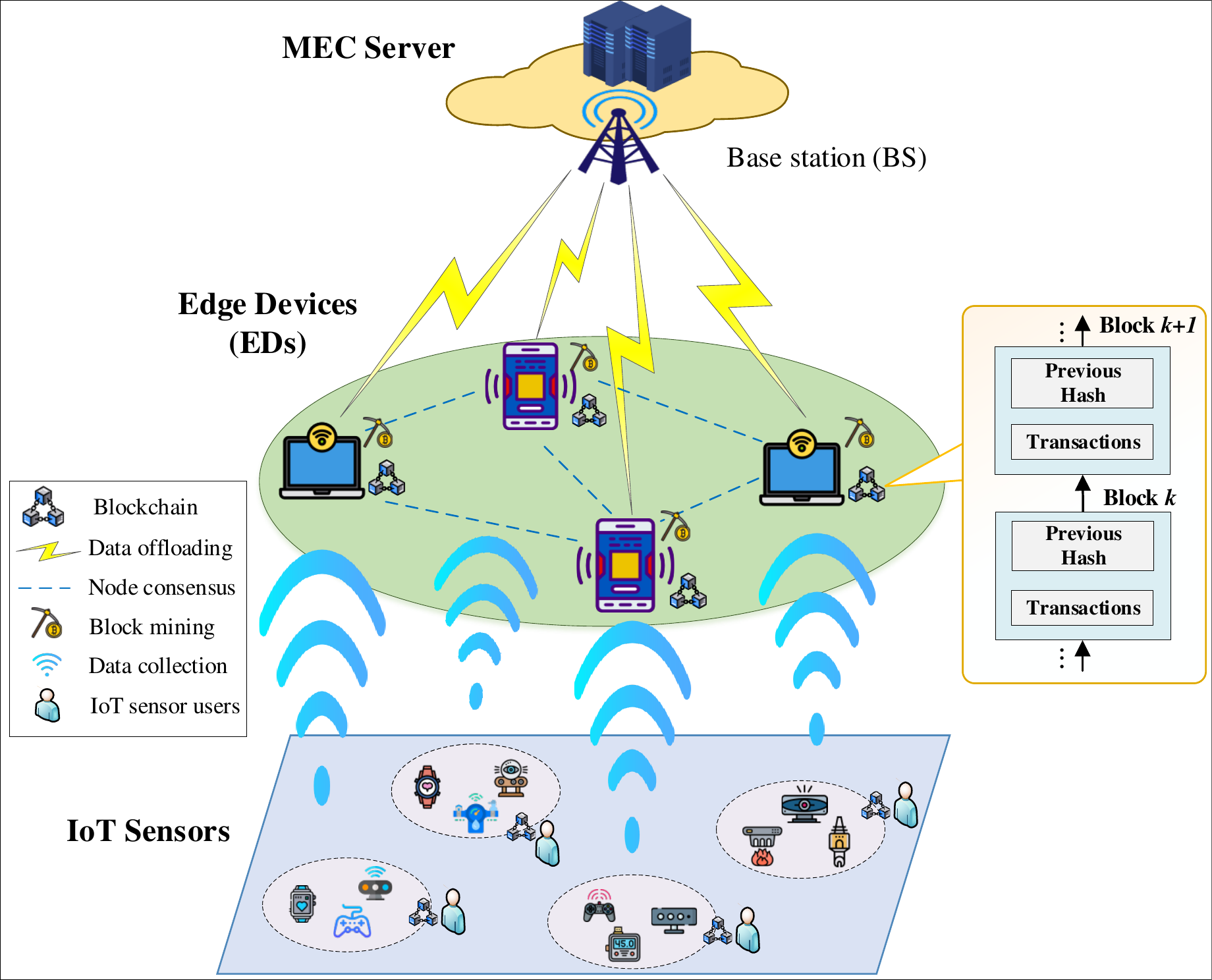} 
	\caption{\textcolor{black}{The proposed cooperative  task offloading and block mining architecture in the B-MEC system. }}
	\label{Overview}
	\vspace{-0.1in}
\end{figure}

{In this context, how to ensure high performance, e.g., system utility, for the B-MEC system is a critical challenge. The task offloading process between EDs and the MEC server consumes much energy and latency, while the operation of blockchain results in delays in the offloading due to mining task execution, which would degrade the overall system utility. Hence, it is paramount to simultaneously consider both task offloading and blockchain mining via a joint design and optimization solution, aiming to enhance the system utility of the B-MEC system.}
\subsection{{Existing Solutions for Intelligent Performance Optimization in B-MEC} }
{To achieve intelligent performance optimization in B-MEC systems, different solutions have been proposed in the open literature.} The authors in \cite{4} considered a blockchain-empowered computation offloading scheme where smart devices can offload their computing tasks to the MEC server under the control of blockchain mining for data integrity. Another work in \cite{5} suggested an online computation offloading approach for both data processing and mining tasks in blockchain-empowered MEC with a deep reinforcement learning (DRL) algorithm {\cite{add2}}, \cite{xiao2020reinforcement}. An intelligent offloading framework with actor-critic DRL was also proposed in \cite{6}, while the study in \cite{7} focused on joint optimization of computation offloading and resource allocation using a double-dueling deep Q-network (DQN) for blockchain-enabled MEC systems.  \cite{8} developed a computation offloading framework for blockchain-based IoT networks with a multi-agent DRL algorithm \cite{add3}.

Despite these research efforts, there are still several urgent issues to be addressed:
\begin{itemize}
	\item \textit{Non-cooperative Offloading:} Most of the existing B-MEC schemes use traditional single-agent DRL algorithms \cite{4,5,6} which \textcolor{black}{exhibit} critical design challenges caused by diversified learning environments. Indeed, each agent only observes its local information in the training without updating the policies of other agents, which makes the learning environment nonstationary \cite{lowe2017multi}. Moreover, non-cooperative multi-agent DRL solutions \cite{7} are not able to learn the mutual offloading policy, which limits the resource utilization for task offloading. 
	\item	\textit{High Blockchain Latency:} The integration of blockchain in MEC potentially results in unnecessary network latency due to block mining \cite{6,7} (i.e., block verification and consensus) which can degrade the overall performance of the B-MEC system. 
	\item	\textit{Lack of Joint Offloading and Mining Design:} In most current B-MEC schemes \cite{6,7,8}, the design and optimization of task offloading and blockchain mining are done separately, leading to a suboptimal performance. To improve {the overall performance of the B-MEC system,} a joint design of task offloading and blockchain mining is highly needed. 
\end{itemize}
\subsection{Our Key Contributions}
Motivated by the aforementioned limitations, we propose a novel cooperative DRL solution for joint task offloading and blockchain mining (TOBM), aiming to maximize the overall B-MEC system utility as a sum of the offloading utility and the mining utility. Our main contributions are highlighted as follows:
\begin{itemize}
	\item 	We propose a novel cooperative TOBM scheme for B-MEC to enable a joint design of task offloading and blockchain mining in Section II. To reduce the blockchain network latency, we develop a new Proof-of-Reputation (PoR) mining mechanism via a lightweight block verification solution.
	\item	We propose a novel cooperative DRL solution in Section III using a multi-agent deep deterministic policy gradient (MA-DDPG) approach \cite{9} to optimize the system utility. 
	\item	{We implement simulations to verify the effectiveness of our proposal in Section IV. We highlight the technical challenges in B-MEC research and discuss several directions for future works in Section V. }
\end{itemize}
  
\section{{Blockchain-Empowered MEC System}}
\subsection{Overview of Network Architecture}
\textcolor{black}{We consider a {cooperative TOBM architecture} in a B-MEC system, as illustrated in Fig.~\ref{Overview}. {An MEC server co-located at a BS provides} computation services for EDs. We assume that each ED has an IoT data task to be executed locally or offloaded to the MEC server. Furthermore, each ED participates in the block mining by using a PoR consensus mechanism.} The key network components of the B-MEC system are described as follows.

\begin{itemize}
	\item \textit{IoT Sensors:} IoT sensors such as cameras, smart meters, and wearables are responsible for sensing physical environments and generating data which need to be  processed to serve end users. IoT sensors also act as lightweight blockchain nodes to transmit data to  EDs. 
	\item \textit{Edge Devices:} Each ED such as a laptop or a powerful smartphone manages a group of IoT sensors under its coverage. Based on the QoE requirements, EDs can use their computational capability to process data tasks locally or offload to a nearby MEC server via wireless links. EDs also work as miners to perform block consensus where IoT sensors' users vote to select representative EDs for mining.
	\item \textit{MEC Server:} In our considered B-MEC system, there is \textcolor{black}{a single MEC sever} to handle computationally extensive data tasks offloaded from EDs. By analyzing the task profile such as  task sizes, channel conditions, and available resources, EDs can make offloading decisions so that the MEC server allocates its resources to execute data tasks under QoE requirements. 
	\item \textit{Blockchain:} A blockchain network is deployed over the MEC system where each ED acts as a blockchain miner \cite{7}. In this paper, we pay attention to a PoR mining design to solve blockchain latency issues. The proposed PoR scheme allows EDs to join the block mining with mining utility enhancement which helps improve the overall performance of the B-MEC system.  
\end{itemize}

\subsection{Task Offloading Model}
We consider a B-MEC system with the set of EDs and the available sub-channels of the BS denoted by $\mathcal{N}$ and $\mathcal{K}$, respectively. It is assumed that each ED $n \in \mathcal{N}$ has an IoT data task including input data and required CPU workload to be executed locally or offloaded to the BS via one of the sub-channels $k \in \mathcal{K}$. Here, the offloading policy is scheduled by a binary variable, which equals 1 (offloading to the MEC server) or 0 (local execution). Each ED $n$ makes offloading decisions based on three main factors: task data size, channel condition, and transmit power level. Moreover, {each ED needs to allocate portion of its computation resource to execute the task locally}. Accordingly, we define four policies to schedule the offloading process of each ED, including offloading decision, channel allocation, transmit power, and computation resource allocation. 

\textcolor{black}{We here formulate an offloading utility function from the QoE perspective, which is characterized by the task computation time (including local time and offloading time) and energy consumption (including local energy and offloading energy). We define a QoE-aware offloading utility function $J_n^{off}$ to provide a trade-off between the time and energy consumption of the task offloading compared with the local execution at each ED. The offloading utility function reflects the offloading improvement in QoE over the local execution.} If the offloading computation cost is lower than the local execution cost, the user utility can be positive, implying the user's QoE improvement. However, if offloading too many tasks, EDs may suffer from higher latency due to the traffic congestion, which reduces the user's QoE. As a result, the user offloading utility can be negative. 
\subsection{{Blockchain Mining Protocol}}
\label{Subsection:Mining}
\begin{figure}
	\centering
	\includegraphics [width=0.99\linewidth]{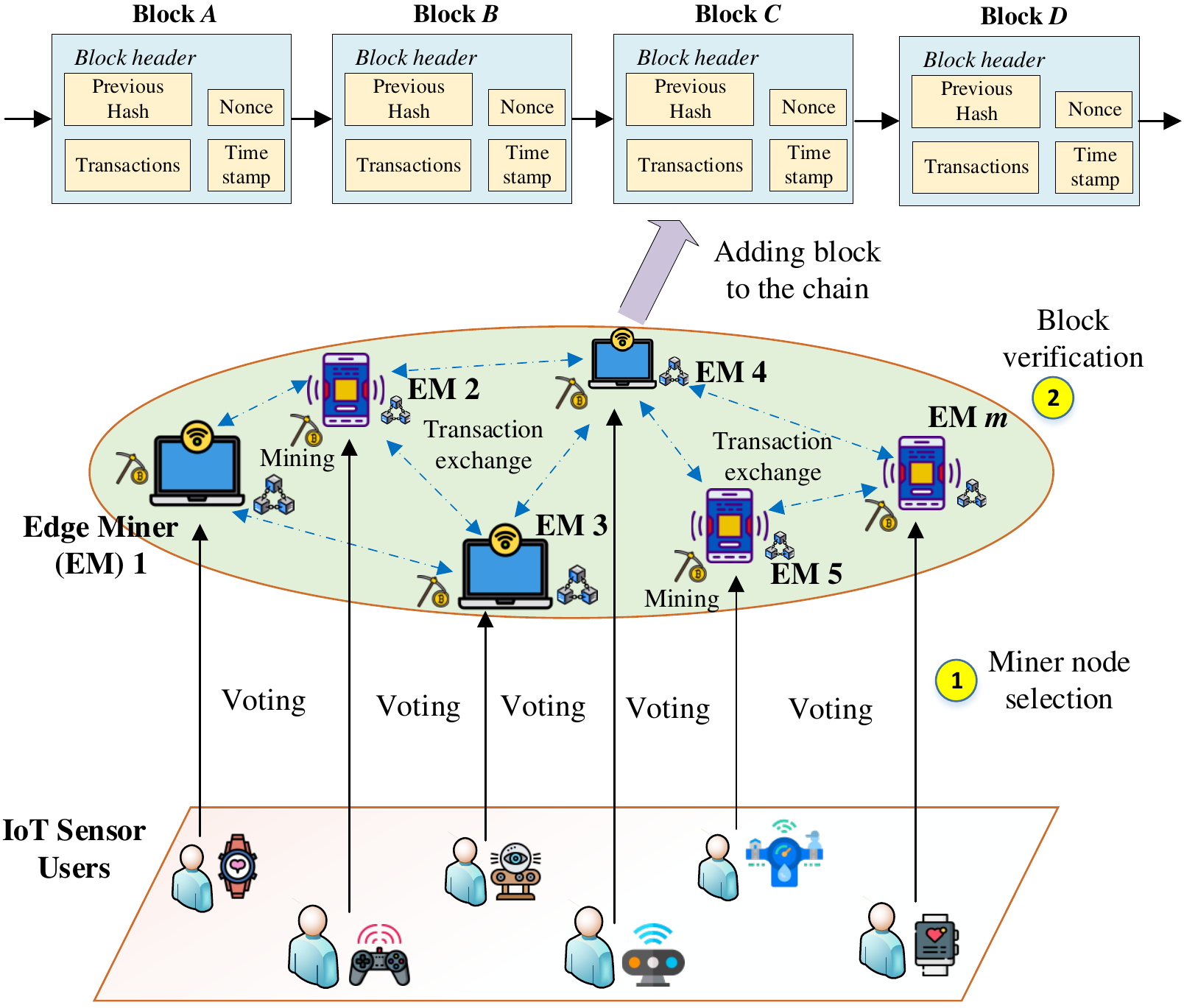}
	\caption{The proposed PoR consensus in our B-MEC system. }
	\label{Fig:PoR}
	\vspace{-0.1in}
\end{figure}
\begin{figure*}[t!]
	\centering
	\begin{subfigure}[t]{0.5\textwidth}
		\centering
		\includegraphics[width=0.99\linewidth]{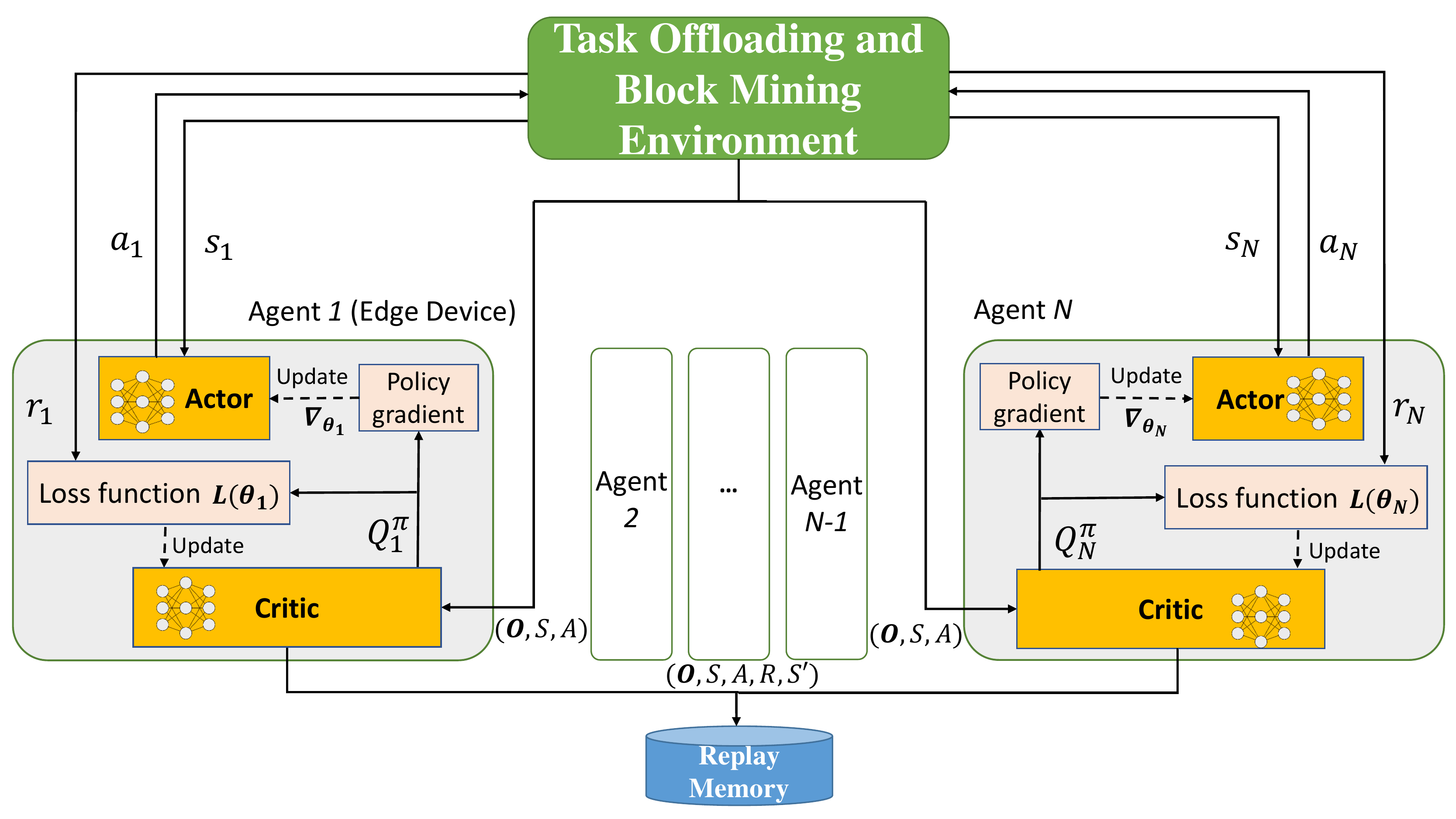}
		\caption{{The proposed MA-DDPG architecture. }}
	\end{subfigure}%
	~
	\begin{subfigure}[t]{0.5\textwidth}
		\centering
		\includegraphics[width=0.99\linewidth]{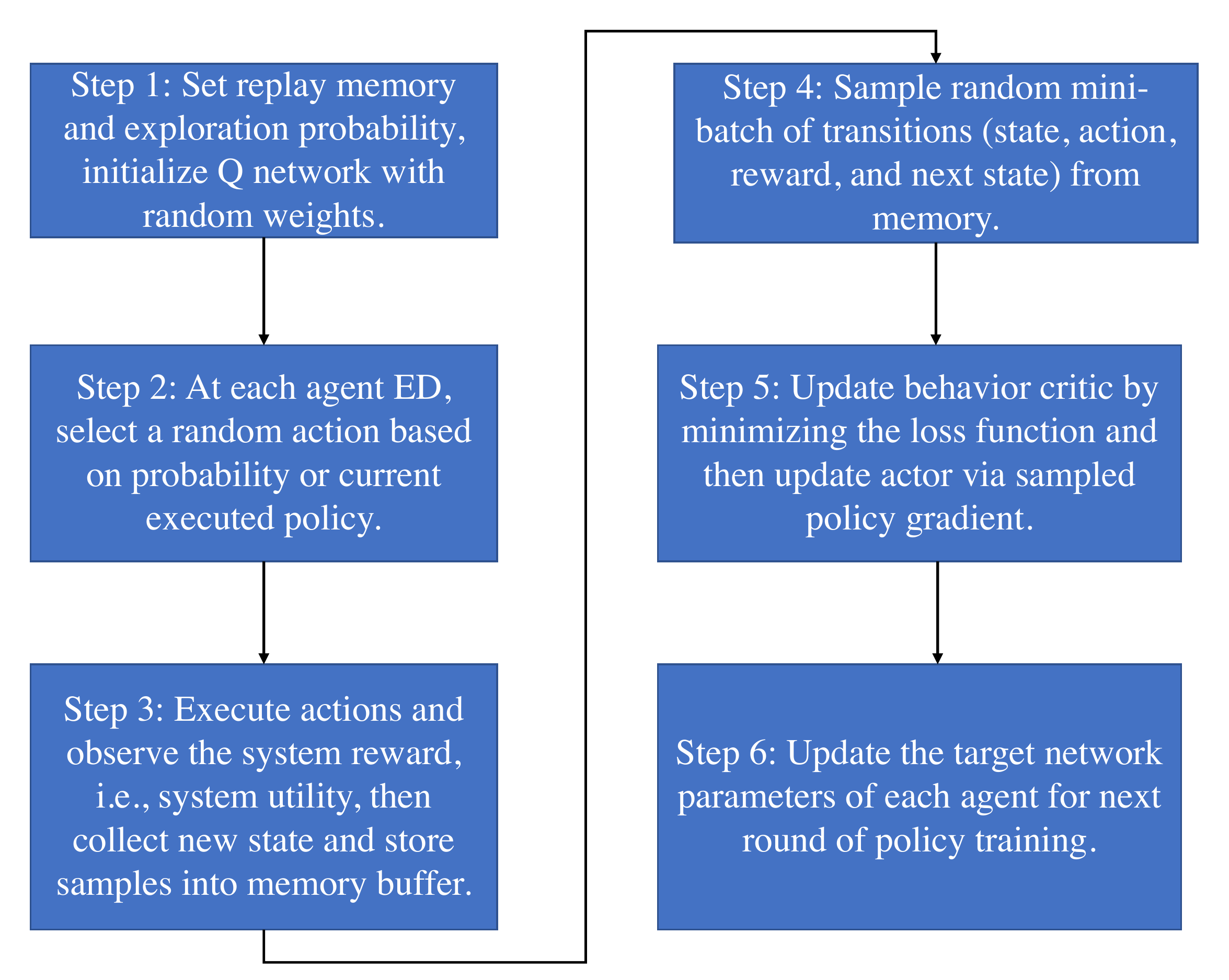} 
	    \caption{\textcolor{black}{The MA-DDPG training procedure. }}
	\end{subfigure}%
	\caption{The design of learning framework for the TOBM system. }
	\label{MADRL}
	\vspace{-0.1in}
\end{figure*}

In the B-MEC system, a crucial component is blockchain consensus that aims to mine the blocks of transactions (i.e., IoT data records) and add them to the blockchain. In traditional consensus mechanisms, e.g., DPoS \cite{add1},\cite{7}, each miner node must implement a repeated verification process across the miner network, which results in unnecessary blockchain latency. Therefore, we here propose a new PoR consensus to solve mining latency issues, including two main parts: miner node formulation and block verification, as illustrated in Fig.~\ref{Fig:PoR}. 
\subsubsection{Miner Node Formulation}
In our B-MEC system, IoT sensors' users participate in the delegate selection process to vote the mining candidates among EDs. \textcolor{black}{In this regard, each IoT user votes for its preferred ED with the most reputation based on its mining latency. Specifically, an ED that exhibits a lower mining latency will have a better reputation via a predefined mining utility function. Based on the calculated reputation score, each IoT user votes for ED candidates based on their reputation ranking.} The top EDs with highest reputation scores are selected to become miners to perform consensus. During its time slot of the consensus process, each miner acts as a block manager which is responsible for performing block generation, verification, and aggregating blocks after being verified. 
\subsubsection{Lightweight Block Verification}
The block manager first produces an unverified block that contains transactions collected by EDs in a given time. Then, the manager broadcasts this block to all other miners within the miner network for verification. Different from the traditional DPoS scheme which relies on a repeated verification process among miners, here we implement a lightweight verification solution. That is, a miner only verifies once with another node during the consensus process, which significantly reduces the verification latency. \textcolor{black}{Specifically, the block manager first divides the block into a set of equal transaction parts that are assigned to each miner within the miner group along with a unique random number. Next, the miner chooses to associate with one of the miners within its group to implement the verification for its assigned transaction part by allocating its CPU resource. If 51\% of miners respond with positive verification, and the sum of random numbers calculated by all miners is equal to a pre-defined number, the block manager accepts the verified block and adds it to blockchain. In summary, the mining procedure includes four stages:  (1) transmitting unverified block from the block manager to the EMs, (2) verifying the local block at each EM, (3) sharing the verification result among two EMs, and (4) transmitting the verification result back to the manager. }



\textcolor{black}{To this end, we build a mining utility function to characterize the mining efficiency of the proposed scheme. Motivated by \cite{10}, we characterize the mining utility $J_n^{mine}$ of each ED $n$ via an exponential function where the mining utility is inversely proportional to the mining latency. Accordingly, an ED that exhibits a lower mining latency has a better mining utility with respect to a CPU resource allocation policy.}  

\section{{Proposed Cooperative DRL Solution for System Utility Optimization}}

\subsection{{System Utility Formulation}}
Here, we formulate the system utility \textcolor{black}{for the proposed TOBM scheme} by taking both offloading utility and mining utility into account. {As explained previously,} the offloading utility $J_n^{off}$ reflects the efficiency of task offloading over local execution from the QoE perspective. Meanwhile, the mining utility $J_n^{mine}$ reflects the efficiency of mining blocks in the B-MEC network via the mining latency metric. Therefore, our key objective is to maximize the total system utility, i.e., $J$, as the sum of the offloading utility $J_n^{off}$ and the mining utility $J_n^{mine}$ of all EDs $n \in \mathcal{N}$, with respect to the offloading policies including offloading decision, channel allocation, transmit power, computation resource allocation, and the mining policy with CPU resource allocation. 

To apply DRL to \textcolor{black}{the formulated TOBM problem}, we need to convert the objective function from a system utility maximization problem to a reward maximization problem. To do this, we formulate the proposed problem using a multi-agent version of the Markov decision process, also known as a Markov game. This is represented by a tuple of agent set $\mathcal{N}$, state set $\mathcal{S}$, action set $\mathcal{A}$, and observation set $\mathcal{O}$ of all agents \cite{add3}. \textcolor{black}{In fact, at each time slot, an MD does not always have access to all states of the environment, but only observes certain states, which is called observations. While states are complete and detailed information that are relevant to the current task (e.g., which specific wireless channel is occupied by a certain MD for offloading), observations are general information received by an MD (e.g., how many wireless channels are occupied).} 

Each ED $n$ is considered as an intelligent agent to learn its optimal policy by observing the local environment formed by the cooperation of EDs and the MEC server, as shown in Fig.~\ref{MADRL}(a). We assume that the considered collaborative task offloading and block mining scheme operates on discrete-time horizon with each time slot $t$ equal and non-overlapping, and the communication parameters remain unchanged during each time slot. Now, we define each item in the tuple at each time slot $t$ as follows:
\begin{itemize}
	\item \textbf{\textit{State:}} The environment state $\mathcal{S}(t)$ at time slot $t$ in \textcolor{black}{the proposed TOBM scheme} includes five components: task state, channel state, power state, resource state, and transaction state. Here, the task state is defined as the matrix of input data and required CPU workload of all EDs. The channel state is defined via a matrix of channel condition variables of all BSs at EDs, where each variable equals 1 (occupied channel) or 0 (available channel). The power state consists of transmit power levels of EDs in each sub-channel. Moreover, the resource state contains the states of available computation resource for data task execution, and transaction state includes transaction data size.
	
	\item \textbf{\textit{Action:}} By observing the environment states, each ED takes an action \textcolor{black}{according to the} offloading decision, channel selection, transmit power selection, computation resource allocation, and CPU resource allocation, to complete task execution and block mining at each time slot $t$. Therefore, the action space of each ED is the combination of the above action sets. Accordingly, the action space $\mathcal{A}(t)$ of the cooperative game can be defined as a matrix of action sets of all agents.
	
	\item \textbf{\textit{System Reward Function:}} The system reward at one time slot $t$ is the sum of the rewards of all EDs. The objective of \textcolor{black}{our formulated TOBM problem} is to maximize the overall system utility $J$ as the sum of the offloading utility and the mining utility. Therefore, we define $J$ as our system reward function.
\end{itemize}
\begin{figure*}[t!]
	\centering
	\begin{subfigure}[t]{0.33\textwidth}
		\centering
		\includegraphics[width=0.99\linewidth]{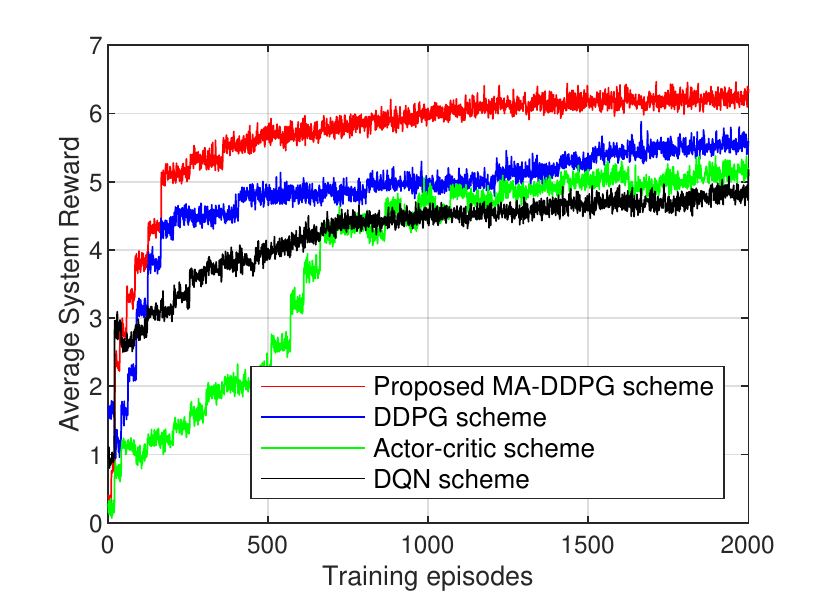} 
		\caption{Average system rewards with different algorithms.   }
	\end{subfigure}
	~
	\begin{subfigure}[t]{0.32\textwidth}
		\centering
		\includegraphics[width=0.99\linewidth]{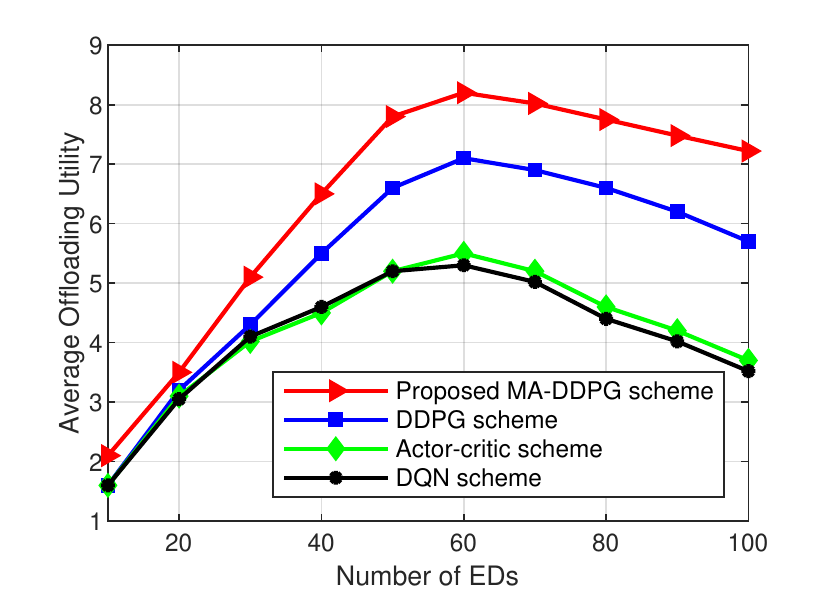} 
		\caption{Average offloading utility with different numbers of EDs. }
	\end{subfigure}%
	~
	\begin{subfigure}[t]{0.32\textwidth}
		\centering
		\includegraphics[width=0.99\linewidth]{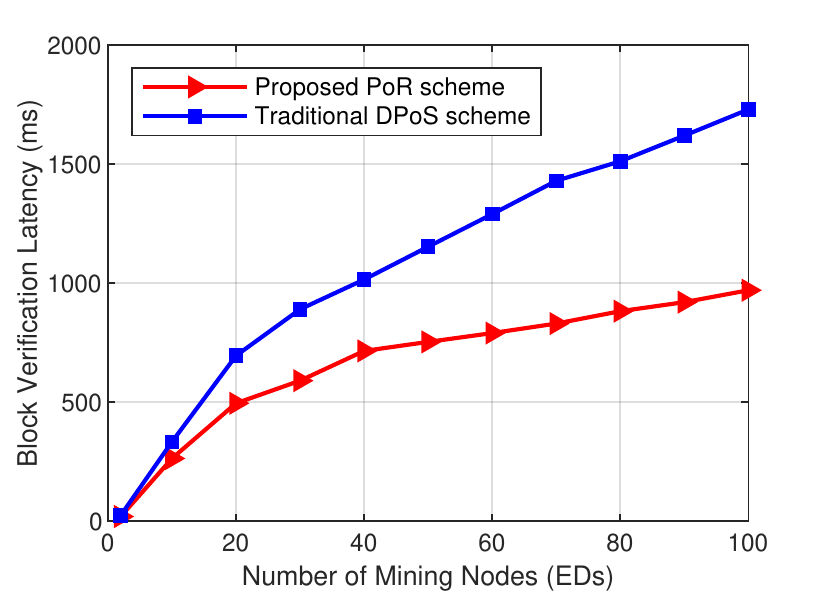} 
		\caption{Comparison of block verification latency. }
	\end{subfigure}%
	\caption{Evaluation of system reward, task offloading, and blockchain performance.}
	\label{Offloading_Performance}
	\vspace{-0.1in}
\end{figure*}

\subsection{Proposed Cooperative DRL Algorithm}

In the literature, most existing schemes have focused on a conventional single-agent \cite{4,5,6} or an independent multi-agent \cite{add3} setting; however, these solutions are unable to obtain the cooperative policies of EDs because of nonstationary and partially observable environments. Indeed, when the policies of other agents change due to computation mode preference, the ED's observation can be changed (nonstationary) Moreover, in independent multi-agent learning schemes, an agent only has the local information and cannot know the updates from other agents due to non-collaboration. This makes the agents' learning algorithm hard to ensure stability and convergence \cite{lowe2017multi}. Thus, we propose a novel cooperative multi-agent DRL scheme using MA-DDPG \cite{9} for \textcolor{black}{our proposed TOBM scheme}. \textcolor{black}{The key reason behind the adaption of DDPG to optimize the system utility is that with this scheme, the actor can directly map states to actions instead of outputting the probability distribution across a discrete action space like DQN, which greatly reduces action sampling complexity. Moreover, given the stochasticity of the policy, there exists high variance of the obtained system reward between different training episodes, where DDPG can come as an efficient solution, by enabling off-policy learning via the joint use of behavior network and target network.}

The MA-DDPG algorithm employs a deep neural network (DNN) as the non-linear approximator to obtain the optimal policies for agents. Each agent updates its parameters to obtain a optimal policy for maximizing its utility. MA-DDPG is a policy gradient-based off-policy \textit{actor-critic} method where each agent comprises the \textit{actor} to make decisions over time slots with a behavior network and \textcolor{black}{the \textit{critic}} to evaluate the behavior of the \textit{actor}, which helps improve its performance. Specifically,  given an episode sample from the memory buffer, the \textit{actor} at each agent updates the behavior network by computing its gradient based on a centralized action-value Q-function. Moreover, the \textit{critic} updates the behavior Q-function for the state-action pair of the \textit{actor} network by minimizing the loss function with the inputs including both local agent's observation and the observations of all other agents. \textcolor{black}{The cooperation of multi-agents helps improve robustness against malicious attacks since this allows multiple agents to monitor the shared policy update via observations \cite{xumisspoke2022}. }

To reduce the computational complexity caused by online training at EDs and solve the nonstationary issues from the concurrently learning process of all EDs, \textcolor{black}{we adopt a centralized learning and decentralized execution solution. In the centralized training step, the information of state-action of all EDs is aggregated by the MEC server to train the DRL model, where each agent can achieve the global view of the learning environment to obtain the observations of other agents for building the collaborative offloading policy.} \textcolor{black}{After training at the MEC server, the learned parameters (i.e. neural weights of DNNs of the actor and critic) are downloaded to each ED to execute the model for decision making based on its own locally observed information. Specifically, given the downloaded neural weights, each agent can easily compute the policy via its DNN with system utility used as the objective function, and then sample an action. Subsequently, each agent executes the sampled action, i.e. making decision to offload the data or not, in the defined TOBM environment given its states to obtain a reward.} The proposed algorithm is illustrated in Fig.~\ref{MADRL}(b). 


\section{Performance Evaluation}
In this section, we conduct numerical simulations to evaluate the performance of \textcolor{black}{the proposed TOBM scheme} in a B-MEC system, employing the widely used Shanghai Telecom dataset\footnote{http://www.sguangwang.com/dataset/telecom.zip}. Here, we consider an MEC network with an MEC server and a maximum of 500 mobile phones as EDs distributed over a \textcolor{black}{1~km~$\times$~1~km} area in Shanghai city. \textcolor{black}{The number of channel sub-bands of the BS is set to 30, and each ED has task CPU workloads of [0.8-1.5]~Gcyles and transmit power range of [0-24]~dBm.} A DNN structure with three hidden layers (64, 32 and 32 neurons) \cite{7} is employed with the Adam optimizer for learning simulation. To prove the advantages of the proposed cooperative MA-DDPG scheme, we compare its performance with the state-of-the-art non-cooperative schemes, including DDPG, actor-critic \cite{6} and DQN \cite{add3}.

\textit{Evaluation of Training Performance:}   Fig.~\ref{Offloading_Performance}(a) shows the learning curves of the average system reward with the increase of learning episodes for the B-MEC system with 50 EDs. It is clear that our MA-DDPG scheme is more robust and yields the best performance in terms of average system reward, compared to baseline schemes. This is because the proposed scheme allows EDs to learn mutually the cooperative  offloading policy which helps reduce the channel congestion and user interference, and enhance computation resource efficiency.  Meanwhile, in the DQN and actor-critic schemes, EDs greedily access the wireless channel spectrum to maximize their own utility without collaboration, which increases the possibility of channel collision and thus results in higher offloading latency. Also, DDPG scheme still remains a non-stationary learning issue and its average reward is lower than that of the MA-DDPG scheme.
\begin{figure*}[t!]
	\centering
	\begin{subfigure}[t]{0.5\textwidth}
		\centering
		\includegraphics[width=0.99\linewidth]{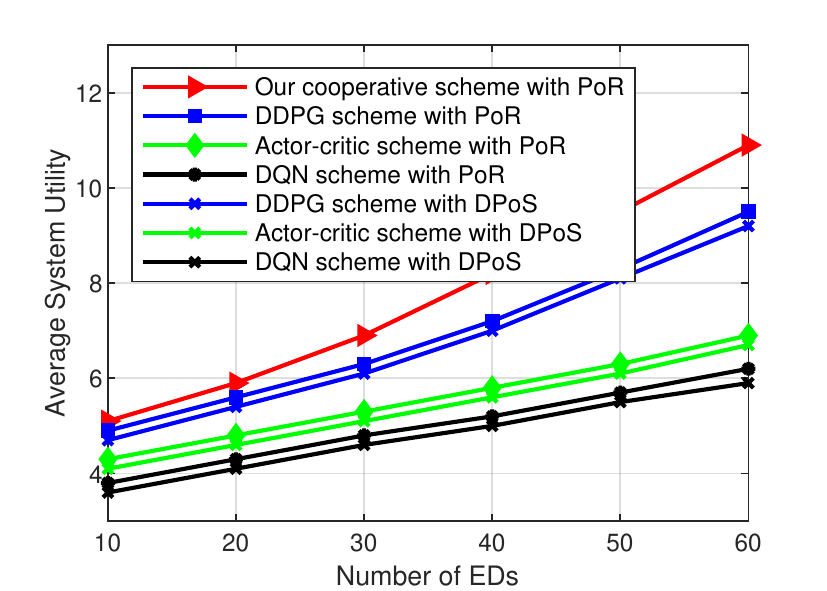} 
		\caption{{Comparison of system utility with non-cooperative schemes. }}
	\end{subfigure}%
	~
	\begin{subfigure}[t]{0.5\textwidth}
		\centering
		\includegraphics[width=0.99\linewidth]{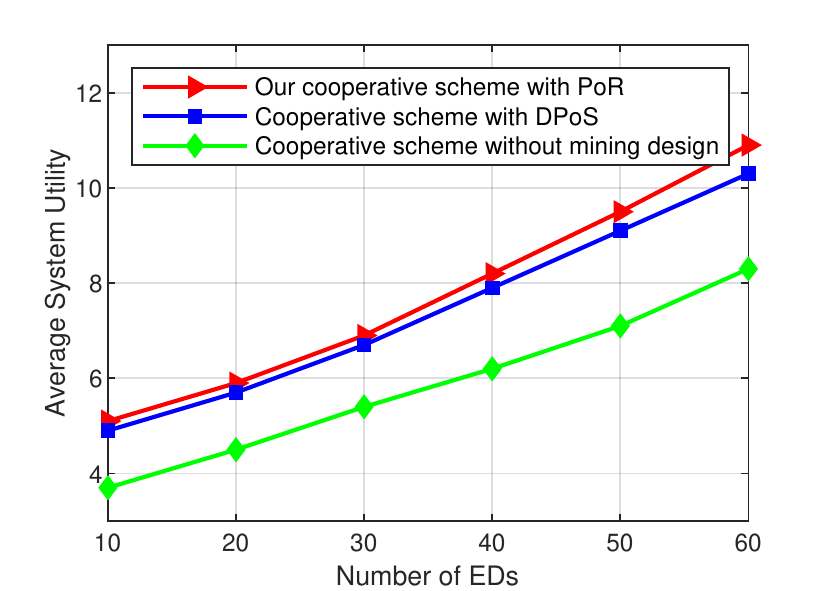} 
		\caption{{Comparison of system utility with cooperative schemes.  }}
	\end{subfigure}%
	\caption{Evaluation of overall system utility performance.}
	\label{Fig:SystemUtility}
	\vspace{-0.1in}
\end{figure*}

\textit{Evaluation of Task Offloading and Blockchain Performance:} Fig.~\ref{Offloading_Performance}(b) indicates the performance of the average offloading utility versus different numbers of EDs. {As can be seen} when the number of EDs is small, the average offloading utility increases with the number of EDs because in this case, the MEC system can support sufficient spectrum and computing resources for handling all tasks of EDs. However, {when exceeding a certain threshold} (i.e., 60 EDs), the offloading utility decreases because the higher the number of offloaded EDs, the higher the competition of resource usage (i.e., channel spectrum). {This in turn increases} the offloading latency, and thus degrades the overall offloading utility. Nevertheless, our MA-DDPG scheme still achieves the best utility performance due to its collaborative offloading policies among EDs compared to other schemes with selfish learning. 


Next, we evaluate our proposed PoR consensus scheme and compare it with the traditional DPoS scheme \cite{7} via the verification block latency metric. We set up 10 transactions per block and vary the numbers of mining nodes from 2 to 200. As shown in Fig.~\ref{Offloading_Performance}(c), our proposed PoR scheme requires significantly less time for mining blocks compared to DPoS, due to the optimized block verification procedure. {Although the block verification latency increases with \textcolor{black}{increasing the number of miners}, our scheme still achieves much better performance than DPoS, which verifies the effectiveness of our lightweight blockchain consensus design.}

\textit{Evaluation of the Overall System Utility Performance:} 
We evaluate the performance of \textcolor{black}{our proposed TOBM scheme} in terms of the overall system utility as the sum of offloading utility and mining utility. The performances of \textcolor{black}{our cooperative TOBM scheme} with our PoR mining design and other non-cooperative schemes with PoR and DPoS mining are illustrated in Fig.~\ref{Fig:SystemUtility}(a). {Unsurprisingly, our cooperative scheme with a PoR mining design achieves the best overall system utility. The reasons for this observation are two-fold.} First, our offloading scheme with a cooperative MA-DDPG algorithm outperforms other non-cooperative offloading schemes in terms of a better offloading utility, {as evidenced in} Fig.~\ref{Offloading_Performance}(a). Second, our PoR design yields a lower mining latency which consequently improves the mining utility. Moreover, due to better mining utility, our PoR design contributes to better overall system utilities in each non-cooperative offloading scheme, compared to the use of DPoS design.

\textcolor{black}{We compare the system utility performance of our proposed TOBM scheme with the cooperative scheme without mining design \cite{6} and the cooperative scheme with DPoS design \cite{7}. As shown in Fig.~\ref{Fig:SystemUtility}(b), our TOBM scheme with PoR design achieves higher system utility than the cooperative scheme with DPoS design, thanks to the better mining utility of our proposed PoR framework. Moreover, compared with our approach, the cooperative scheme in \cite{6} has the lowest system utility due to the lack of mining design. This simulation result also reveals that a joint design of offloading and mining is of paramount importance to improving the overall system performance in B-MEC systems.}

\begin{figure*}[t!]
	\centering
	\begin{subfigure}[t]{0.5\textwidth}
		\centering
		\includegraphics[width=0.99\linewidth]{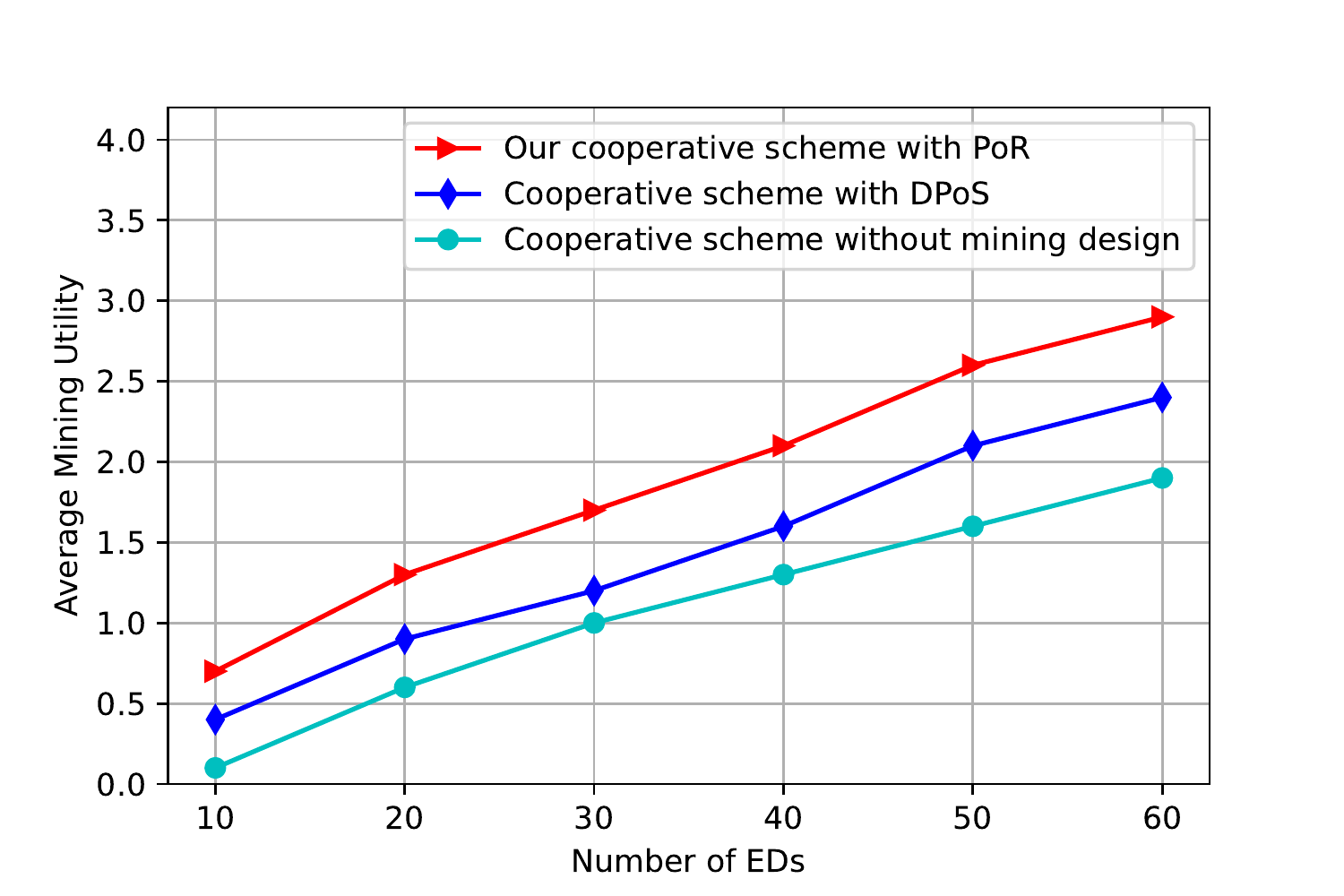} 
		\caption{{Comparison of mining utility with cooperative schemes.  }}
		
	\end{subfigure}%
	~
	\begin{subfigure}[t]{0.5\textwidth}
		\centering
		\includegraphics[width=0.99\linewidth]{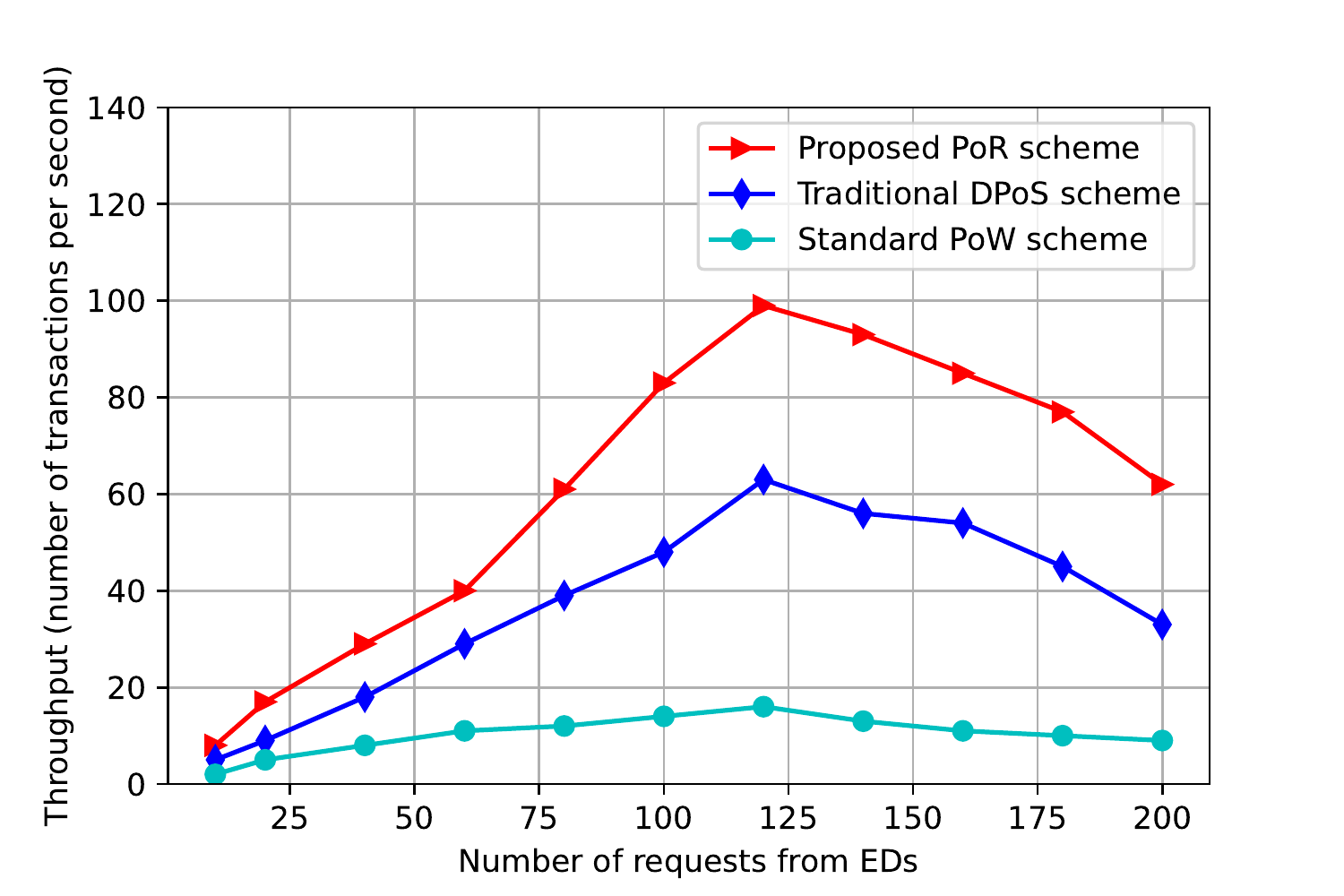} 
		\caption{{Comparison of blockchain throughput with mining schemes. }}
	\end{subfigure}%
	\caption{\textcolor{black}{Evaluation of blockchain performance.}}
	\label{Fig:blockchainper}
	\vspace{-0.1in}
\end{figure*}

\textcolor{black}{We compare the mining utility of our scheme with other cooperative schemes in Fig.~\ref{Fig:blockchainper}(a). Due to a lower mining latency achieved by the proposed consensus design, our scheme yields a better utility compared with its counterparts. The traditional approach without mining design has the highest mining latency, resulting in the lowest utility. Moreover, we investigate the throughput in Fig.~\ref{Fig:blockchainper}(b) which is defined as the number of successful transactions per second. We set five transactions per block and make offloading requests ranging from 10 to 200. Compared with DPoS and Proof-of-Work (PoW) schemes, the fast block verification rate of our scheme significantly enhances the throughput, before its performance decreases when the number of requests is higher than 120 since the system cannot handle excessive requests under this configuration.
}

{\color{black}\section{Research Challenges and Future Perspectives}

\subsection{User Mobility} In realistic B-MEC systems, EDs can move with high speed in wireless networks (e.g., in on-vehicle applications). {This makes their location highly dynamic which has direct impacts on offloading decision making. For example, an ED is likely to execute locally if it moves out of BS coverage, which makes the offloading design ineffective.} Therefore, the user mobility needs to be considered in the offloading formulation in future B-MEC systems, where dynamic characteristics such as {velocity and channel conditions should be taken into account to come up with a robust mobility-aware offloading policy}. 
\subsection{Data Privacy}
In this work, {the DRL training performed at the MEC server potentially raises data privacy leakage due to the data exchange during the training.} Federated learning \cite{3} can be an attractive solution to perform collaborative training, where only trained parameters are shared with the MEC server while actual data and user information are stored at local EDs for privacy enhancement. In the B-MEC context, for example, each ED can run a DRL function to learn its offloading and mining policy based on its own local observation. Then, the participating EDs can communicate with the MEC server for DRL model aggregation (e.g., model averaging) to create a new global DRL model without sharing their private data. 

\subsection{{Incentive Issues}} \textcolor{black}{In practice, how to encourage EDs, which serve as both offloading and mining nodes, to join computation and data mining for the long term is a critical challenge for B-MEC systems. Although blockchain is able to incentivize EDs via coin payment based on their mining effort, it is still not enough to compensate for the energy resources consumed for computation task offloading. Thus, it is important to jointly consider incentives, offloading and mining in B-MEC system optimization. A possible direction is to design a smart contract-inspired incentive mechanism to accelerate the data offloading and block mining. Another interesting area is to jointly optimize utility with respect to monetary benefits and resource usage of offloading and mining. }

\section{Conclusions}
This article proposed the novel concept of TOBM to assist  B-MEC systems. A joint design of offloading and mining was considered, where a PoR consensus mechanism was proposed. Then, a cooperative DRL approach was proposed to solve the TOBM problem, showing a huge system utility improvement over the existing cooperative and non-cooperative schemes. Finally, we highlighted the key research challenges and promising directions for future B-MEC research.


\bibliography{Ref}

\begin{thebibliography}{10}
\providecommand{\url}[1]{#1}
\csname url@samestyle\endcsname
\providecommand{\newblock}{\relax}
\providecommand{\bibinfo}[2]{#2}
\providecommand{\BIBentrySTDinterwordspacing}{\spaceskip=0pt\relax}
\providecommand{\BIBentryALTinterwordstretchfactor}{4}
\providecommand{\BIBentryALTinterwordspacing}{\spaceskip=\fontdimen2\font plus
\BIBentryALTinterwordstretchfactor\fontdimen3\font minus
  \fontdimen4\font\relax}
\providecommand{\BIBforeignlanguage}[2]{{%
\expandafter\ifx\csname l@#1\endcsname\relax
\typeout{** WARNING: IEEEtran.bst: No hyphenation pattern has been}%
\typeout{** loaded for the language `#1'. Using the pattern for}%
\typeout{** the default language instead.}%
\else
\language=\csname l@#1\endcsname
\fi
#2}}
\providecommand{\BIBdecl}{\relax}
\BIBdecl

\bibitem{1}
Y.~Dai, D.~Xu, S.~Maharjan, Z.~Chen, Q.~He, and Y.~Zhang, ``Blockchain and
  {Deep} {Reinforcement} {Learning} {Empowered} {Intelligent} {5G} {Beyond},''
  \emph{IEEE Network}, vol.~33, no.~3, pp. 10--17, May 2019.

\bibitem{3}
H.~Yao, T.~Mai, J.~Wang, Z.~Ji, C.~Jiang, and Y.~Qian, ``{Resource} {Trading}
  in {Blockchain}-based {Industrial} {Internet} of {Things},'' \emph{IEEE
  Transactions on Industrial Informatics}, vol.~15, no.~6, pp. 3602--3609,
  2019.

\bibitem{4}
X.~Xu, X.~Zhang, H.~Gao, Y.~Xue, L.~Qi, and W.~Dou, ``{BeCome}:
  {Blockchain}-{Enabled} {Computation} {Offloading} for {IoT} in {Mobile}
  {Edge} {Computing},'' \emph{IEEE Trans. Industr. Infor.}, vol.~16, no.~6, pp.
  4187--4195, Jun. 2020.

\bibitem{add1}
J.~Kang, Z.~Xiong, D.~Niyato, D.~Ye, D.~I. Kim, and J.~Zhao, ``Toward {Secure}
  {Blockchain}-{Enabled} {Internet} of {Vehicles}: {Optimizing} {Consensus}
  {Management} {Using} {Reputation} and {Contract} {Theory},'' \emph{IEEE
  Trans. Veh. Tech.}, vol.~68, no.~3, pp. 2906--2920, Mar. 2019.

\bibitem{5}
X.~Qiu, L.~Liu, W.~Chen, Z.~Hong, and Z.~Zheng, ``Online {Deep} {Reinforcement}
  {Learning} for {Computation} {Offloading} in {Blockchain}-{Empowered}
  {Mobile} {Edge} {Computing},'' \emph{IEEE Trans. Veh. Tech.}, vol.~68, no.~8,
  pp. 8050--8062, Aug. 2019.

\bibitem{add2}
J.~Wang, C.~Jiang, H.~Zhang, Y.~Ren, K.-C. Chen, and L.~Hanzo, ``Thirty {Years}
  of {Machine} {Learning}: {The} {Road} to {Pareto}-optimal {Wireless}
  {Networks},'' \emph{IEEE Communications Surveys and Tutorials}, vol.~22,
  no.~3, pp. 1472--1514, 2020.

\bibitem{xiao2020reinforcement}
L.~Xiao, Y.~Ding, D.~Jiang, J.~Huang, D.~Wang, J.~Li, and H.~V. Poor, ``A
  {Reinforcement} {Learning} and {Blockchain}-based {Trust} {Mechanism} for
  {Edge} {Networks},'' \emph{IEEE Transactions on Communications}, vol.~68,
  no.~9, pp. 5460--5470, 2020.

\bibitem{6}
J.~Feng, F.~R. Yu, Q.~Pei, X.~Chu, J.~Du, and L.~Zhu, ``Cooperative
  {Computation} {Offloading} and {Resource} {Allocation} for
  {Blockchain}-{Enabled} {Mobile}-{Edge} {Computing}: {A} {Deep}
  {Reinforcement} {Learning} {Approach},'' \emph{IEEE Inter. Things J.},
  vol.~7, no.~7, pp. 6214--6228, Jul. 2020.

\bibitem{7}
F.~Guo, F.~R. Yu, H.~Zhang, H.~Ji, M.~Liu, and V.~C. Leung, ``Adaptive
  {Resource} {Allocation} in {Future} {Wireless} {Networks} {With} {Blockchain}
  and {Mobile} {Edge} {Computing},'' \emph{IEEE Trans. Wireless Commun.},
  vol.~19, no.~3, pp. 1689--1703, Mar. 2020.

\bibitem{8}
Z.~Li, M.~Xu, J.~Nie, J.~Kang, W.~Chen, and S.~Xie, ``{NOMA}-{Enabled}
  {Cooperative} {Computation} {Offloading} for {Blockchain}-{Empowered}
  {Internet} of {Things}: {A} {Learning} {Approach},'' \emph{IEEE Inter. Things
  J.}, pp. 1--1, Aug. 2020.

\bibitem{add3}
J.~Heydari, V.~Ganapathy, and M.~Shah, ``Dynamic {Task} {Offloading} in
  {Multi}-{Agent} {Mobile} {Edge} {Computing} {Networks},'' in \emph{2019 IEEE
  Global Communications Conference (GLOBECOM)}, 2019, pp. 1--6.

\bibitem{lowe2017multi}
R.~Lowe, Y.~Wu, A.~Tamar, J.~Harb, P.~Abbeel, and I.~Mordatch, ``Multi-{Agent}
  {Actor}-{Critic} for {Mixed} {Cooperative}-{Competitive} {Environments},'' in
  \emph{Proc. Advances in {Neural} {Infor.} {Process} {Sys.}}, 2017, pp.
  6379--6390.

\bibitem{9}
D.~Kwon, J.~Jeon, S.~Park, J.~Kim, and S.~Cho, ``Multi-{Agent} {DDPG}-based
  {Deep} {Learning} for {Smart} {Ocean} {Federated} {Learning} {IoT}
  {Networks},'' \emph{IEEE Inter. Things J.}, pp. 1--1, Apr. 2020.

\bibitem{10}
Y.~Dai, D.~Xu, K.~Zhang, S.~Maharjan, and Y.~Zhang, ``Deep {Reinforcement}
  {Learning} and {Permissioned} {Blockchain} for {Content} {Caching} in
  {Vehicular} {Edge} {Computing} and {Networks},'' \emph{IEEE Trans. Veh.
  Tech.}, vol.~69, no.~4, pp. 4312--4324, Apr. 2020.

\bibitem{xumisspoke2022}
W.~Xue, W.~Qiu, B.~An, Z.~Rabinovich, S.~Obraztsova, and C.~K. Yeo, ``Mis-spoke
  or {Mis-lead}: {Achieving} {Robustness} in {Multi}-{Agent} {Communicative}
  {Reinforcement} {Learning},'' Jan. 2022, arXiv: 2108.03803.

\end{thebibliography}
\bibliographystyle{IEEEtran}
\vskip -2\baselineskip 
\begin{IEEEbiographynophoto}{Dinh C. Nguyen}
\balance	is a Postdoctoral Associate at the School of Electrical and Computer Engineering, Purdue University, USA. He obtained the Ph.D. degree at Deakin University, Australia in 2022.  His research focuses on federated learning, blockchain, and edge computing. He is an Associate Editor of the IEEE Open Journal of the Communications Society. 
\end{IEEEbiographynophoto}
\vskip -2\baselineskip 

\begin{IEEEbiographynophoto}{Van-Dinh Nguyen}
 received Ph.D. degree in electronic engineering from Soongsil University, South Korea 2018.  He is currently a Research Associate with the Interdisciplinary Centre for Security, Reliability and Trust (SnT), University of Luxembourg. His current research activity is focused on the mathematical modeling of 5G cellular networks and machine learning for wireless communications. Dr. Nguyen received several best conference paper awards, IEEE Communications Letters Exemplary Editor 2019 and 2021.
\end{IEEEbiographynophoto}
\vskip -2\baselineskip 
\begin{IEEEbiographynophoto}{Ming Ding}
	is currently a Senior Research Scientist with the CSIRO Data61, Sydney, NSW, Australia. His research interests include information technology, data privacy and security, machine learning and AI. He has authored over 100 articles in IEEE journals and conferences. He is an Editor of the IEEE Transactions on Wireless Communications and the IEEE Wireless Communications Letters. 
\end{IEEEbiographynophoto}
\vskip -2\baselineskip 
\begin{IEEEbiographynophoto}{Symeon Chatzinotas}
 is Full Professor/Chief Scientist I and Co-Head of the SIGCOM Research Group at SnT, University of Luxembourg. He was a co-recipient of the 2014 IEEE Distinguished Contributions to Satellite Communications Award,  the CROWNCOM 2015 Best Paper Award and the 2018 EURASIC JWCN Best Paper Award. He has (co-)authored more than 400 technical papers in refereed international journals, conferences and scientific books. He is currently in the editorial board of the IEEE Open Journal of Vehicular Technology and the International Journal of Satellite Communications and Networking.
\end{IEEEbiographynophoto}
\vskip -2\baselineskip 

\begin{IEEEbiographynophoto}{Pubudu N. Pathirana}
	is a full Professor and the Director of Networked Sensing and Control group at the School of Engineering, Deakin University, Geelong, Australia. He was a visiting professor at Yale University in 2009. His current research interests include Bio-Medical assistive device design, mobile/wireless networks, and rehabilitation robotics.
\end{IEEEbiographynophoto}
\vskip -2\baselineskip 
\begin{IEEEbiographynophoto}{Aruna Seneviratne}
	is currently a Foundation Professor of telecommunications with the University of New South Wales, Australia, where he holds the Mahanakorn Chair of telecommunications. His current research interests are in physical analytics: technologies that enable applications to interact intelligently and securely with their environment in real time. 
\end{IEEEbiographynophoto}
\vskip -2\baselineskip 
\begin{IEEEbiographynophoto}{Octavia A Dobre}
	 is a Professor and Research Chair at Memorial University, Canada. Her research interests include technologies for beyond 5G, as well as optical and underwater communications. She published over 300 referred papers in these areas. Dr. Dobre serves as the Editor-in-Chief (EiC) of the IEEE Open Journal of the Communications Society. She was the EiC of the IEEE Communications Letters, a senior editor and an editor with prestigious journals, as well as General Chair and Technical Co-Chair of flagship conferences in her area of expertise. She is Fellow of the IEEE and the Engineering Institute of Canada.
\end{IEEEbiographynophoto}
\vskip -2\baselineskip 
\begin{IEEEbiographynophoto}{Albert Y. Zomaya}
	is Peter Nicol Russell Chair Professor of Computer Science and Director of the Centre for Distributed and High Performance Computing at Sydney University. He has published more than 600 scientific papers and is an author, co-author, or editor of more than 30 books. He is the Editor-in-Chief of the ACM Computing Surveys and served in the past as the Editor-in-Chief of the IEEE Transactions on Computers and the IEEE Transactions on Sustainable Computing. He is a Fellow of the AAAS, the IEEE, the IET (UK), and was inducted in 2022 into the Australian Academy of Science.
\end{IEEEbiographynophoto}
\end{document}